\documentclass[12pt]{article}
\usepackage{amsmath,amsfonts,amssymb,latexsym,graphicx}
\DeclareGraphicsExtensions{.pdf,.jpg}

\newtheorem{theorem}{Theorem}[section]

\numberwithin{equation}{section}
\def\be{\begin{equation}}
\def\ee{\end{equation}}
\def\bq{\begin{eqnarray}}
\def\eq{\end{eqnarray}}
\def\beq{\begin{eqnarray}}
\def\eeq{\end{eqnarray}}

\def\beq{\begin{equation}}
\def\eeq{\end{equation}}

\begin{document}

\title{\Large{\textsc{Infinity in string  cosmology: A review through open problems}}}
\author{{\large\textsc{Ignatios Antoniadis$^{1,2}$\thanks{\texttt{antoniad@lpthe.jussieu.fr}}, Spiros Cotsakis\footnote{On leave from the University of the Aegean, 83200 Samos, Greece.}\,\,$^{3}$\thanks{\texttt{skot@aegean.gr}}}}, 
\\[10pt]
$^1$LPTHE, UMR CNRS 7589, Sorbonne Universit\'es, UPMC Paris 6,\\
4 place Jussieu, T13-14, 75005 Paris, France\\
$^2$ Albert Einstein Center for Fundamental Physics, ITP,\\
University of Bern, 
Sidlerstrasse 5 CH-3012 Bern, Switzerland\\
$^{3}$Department of Mathematics, 
American University of the Middle East\\
P. O. Box 220 Dasman, 15453, Kuwait }
\maketitle
\begin{abstract}
\noindent We review recent developments in the field of string cosmology with particular emphasis on open problems having to do mainly with geometric asymptotics and singularities. We discuss outstanding issues in a variety of currently popular themes, such as tree-level string cosmology asymptotics, higher-order string correction effects, M-theory cosmology, braneworlds, and finally ambient cosmology.
\end{abstract}


\section{Introduction}
The observed recession of galaxies  \cite{hubble} and the existence of the cosmic microwave background radiation (CMB) \cite{pw,dicke} motivated the inspiration and development of the standard model of cosmology  \cite{wein1}. According to this theory the universe is well described by a Friedmann-Robertson-Walker geometric dynamics all the way back from the time of electron-positron annihilation (radiation domination period) up to the present.

More recent  measurements of the
microwave background temperature anisotropies in combination with the fact that very distant objects known as supernovae type-Ia appear much fainter than what standard matter-dominated cosmological theory predicts, point to an additional effect, the observed acceleration of the expansion of the universe, and pose the fundamental problem of what
causes this accelerated expansion and adequately explains the observations.
This observed acceleration in the present phase of the evolution of the universe is in obvious compliance with more recent data for the redshift  \cite{iye} and from the Hubble diagram of Type Ia Supernovae \cite{perl,riess}, in conjunction with various  harmonic analyses of the CMB fluctuations and anisotropies \cite{debe,hana,alve} with a very adequate level of statistical precision and confidence \cite{sper1,riess2,sper2}.

These observations also point, with increasing plausibility, to an earlier, vacuum energy dominated period in the history of the universe before the radiation period, where the universe was expanding with the scale factor growing more or less exponentially, a period known as cosmological inflation, cf. \cite{guth,wein1,linde1}. Inflation not only resolves various of the puzzles of the standard model of cosmology but also explains beautifully the origin of the CMB anisotropies, leading to a very persuasive theory of cosmological fluctuations, cf., e.g.,  \cite{Mukhanov,wein1,llyth}.
It is generally accepted  today that the standard model of cosmology together with its inflationary extension describe well the observed properties of the universe and reveal many fundamental aspects of the universe  \cite{wein1}.

The success of the standard and inflationary  cosmology is not accidental but due to its reliance   on general relativity and the quantum theory of fields respectively, perhaps two of the  most successfully established theories of theoretical physics available today \cite{wein2,wein3}. However, and despite all its success, there are two fundamental, unresolved issues associated with standard cosmological theory  lacking a satisfactory explanation within its realm and pointing to conceptual difficulties that lie far beyond it. These two issues cannot be properly addressed either by the standard cosmological model or by any of its inflationary extensions, and prevent the standard, accepted cosmological picture to be the final one, that is a  complete theory that would  eventually offer us a more convincing framework to study the structure and ultimate fate of the universe. These two issues are the cosmological constant problem, and the singularity problem.

The  cosmological constant problem  \cite{wein4} is the question of why the cosmologically observed value of the total, effective vacuum energy,
$\rho_\lambda\sim 10^{-47} GeV^4,$ is so much smaller than the expected quantum field theory estimate of the vacuum energy (empty space),
$\langle\rho\rangle_{vac}\sim 2\times 10^{71} GeV^4.$ In other words, the Einstein cosmological constant, $\lambda$, contributes a term equal to $\lambda /8\pi G$ to $\rho_\lambda$, so that when we solve the equation $\rho_\lambda = \lambda /8\pi G +\langle\rho\rangle_{vac}$, we find the unnaturally
small value $\lambda /8\pi G= 10^{-118}GeV^4$ for the cosmological constant. This is a serious problem for theoretical physics because, since $\rho_\lambda$ comes from general relativity and $\langle\rho\rangle_{vac}$ from quantum field theory, it follows that there is an unexplained discrepancy between the two theories, in a regime where both are perfectly valid,
unless a severe fine-tuning of about 120 decimal places is performed.

The second problem is the so-called singularity, or initial state,  problem. According to the singularity theorems of general relativity, spacetime singularities are a generic prediction of the theory under very plausible assumptions of the causal and matter character (global hyperbolicity, positivity of energy density), usually accompanied with a blow up in the spacetime curvature and the thermodynamical properties of matter  \cite{he}. This means that spacetime must come to an end at generic spacetime singularities and further analyses relating to spacetime structure and evolution beyond such a point cannot really be made.  More recent completeness theorems  \cite{ycb1} need to assume  that various curvatures are bounded to get geodesic completeness, and so cannot alleviate the break down of the theory  predicted by the original singularity theorems. It follows that geodesic incompleteness is tied with a blow up in the curvature. Therefore according to these theorems classical general relativity must loose its predictive power as a physical theory near regimes where the gravitational field is sufficiently strong. (We note that potential-dominated inflation has a similar problem, namely it cannot be past eternal  \cite{borde}, so the inflationary extension of the standard model of cosmology cannot solve the singularity problem.)

This state of affairs opens up a host of new possibilities, theories which may be formulated and studied outside the realm of general relativity  in an arbitrary number of dimensions and have dynamical equations and  matter fields  `arbitrarily' prescribed (or arbitrarily modified with respect to the standard Einstein equations and matter fields). The new problem then becomes to examine what these new models would imply about the ultimate nature, origin and fate of the universe. Indeed, in this way, mathematical models of cosmological importance have in recent decades acquired a sparkling diversity. They may be
constructed using a wide variety of cosmological spacetimes, theories of gravitation and possible matter fields, resulting in an
interesting and seemingly legitimate (at least mathematically) web of theories as the (rather incomplete) Table below shows (cf.  \cite{sc02}).
\begin{center}
\begin{tabular}{l|c|r}
\multicolumn{3}{c}{\emph{\textbf{Cosmologies}}}\cr\hline {\bf
{Theories of gravity}} & {\textbf{Spacetimes}} &
{\textbf{Matterfields}}\cr\hline\hline General Relativity &
de Sitter & Vacuum \cr Higher Derivative  & FRW & Fluids \cr
Scalar-Tensor, Multi-field  & Bianchi,  & Scalar fields\cr Quantum Cosmology &
G\"{o}del & $n$-form fields\cr Varying constants   & Generic & Phantoms,
tachyons \cr\hline
\end{tabular}
\end{center}
In the vast literature of  cosmological models,  one finds  choices from this Table  of the sort GR/generic/vacuum, HD/FRW/vacuum,  ST/Bianchi/fluid,   and so on. However, it is not possible that all models constructed in this more or less \emph{ad hoc} way be a priori physically realistic. In fact, each  choice of a \emph{cosmology} (here meant in a somewhat scholastic way a triplet of the form `Theory of gravity/Spacetime/Matterfield'), would be based on generally inequivalent physical assumptions than any other, so that it would  become impossible for  them to be all viable in the end, irrespectively of their success! To put it in another way, although there is nothing to stop us pursuing any one of these `theories'-choices from the Table above (and similarly of course many others not included here),  a basic unifying principle underlying such an approach is totally absent.

This is precisely where string theory comes in and offers a fresh, new and completely revolutionary  approach to this old subject. For it postulates the missing unifying principle for all fields and interactions, in that every field interacting with the string should be contained in the spectrum of states associated with the quantization of its free oscillations, in such a way as to maintain a certain kind of conformal invariance present in all string models \cite{gsw,pol,zwie}. As such, string theory bears a very remarkable relation to both gravitation and cosmology, and, in particular, to the two problems mentioned above, namely, the cosmological constant problem and the singularity problem.
But equally important string-theoretic considerations are for the issue of the initial state of the universe, where general relativity breaks down and we need a new unifying principle in order to proceed. (Of course string theory has  remarkable connections with particle physics, condensed matter physics and statistical mechanics, but these are unaffected by the arguments of relevance here.)

String gravitation and cosmology  is then a largely unexplored area of theoretical physics of remarkable promise, to which the current review intends to shed new light to a number of basic issues.
In this paper, we aim to address systematically  the important issue of \emph{infinity} inherent in both the cosmological constant problem and the cosmological singularity problem in the unified context of string-inspired models.

It is truly remarkable that in string theory all major aspects in the web of theories freely prescribed as discussed above, reappear  but in a way that is not free anymore. They must be consistent with the rules of string quantization and conformal invariance, and thus are now tightly fixed under these string-theoretic constraints. We therefore expect that these constraints will lead to interesting new effects not only for their initial states, but also for the cosmological dynamics of string models near any regime where the field approaches a `singularity' (an example of the latter situation arises in the so-called `self-tuning' models, see below).

Hence,  in this  paper we will be deeply rooted in string cosmology (see \cite{gasp} for a recent introduction to this general area), and more generally, in string-theoretic models which most naturally emerge when considering:
\begin{itemize}
\item[A.] Tree-level equations both in the string and the Einstein frame
\item[B.] First-order $\alpha'$-corrections
\item[C.] M-theory cosmology
\item[D.] Braneworlds and self-tuning.
\end{itemize}
We  aim to address below in a series of open problems some of the issues of infinity and asymptotic structure   in all four areas of string theory and cosmology referred to above. More specifically, we will present a series of open problems relevant to  the string phase of the models and are related to:
\begin{itemize}
\item[i.] the nature of cosmological `singularities'
\item[ii.] the asymptotic properties of regular solutions, and
\item[iii.] the genericity of the found solutions,
\end{itemize}
Before we proceed further, we make the following remark.  The quotation marks used in the word singularity above, mean to underline the fact that in string-theoretic cosmological considerations the typical general relativistic spacetime singularities are replaced by other regimes. A typical such asymptotic state in string cosmology is the so-called string perturbative vacuum  \cite{gv1}, the exact opposite of a very hot, highly curved and exceedingly dense big bang of the standard model of cosmology. Other string asymptotic regimes correspond to  Planckian or trans-Planckian states which emerge due to the non-perturbative nature of the string models and take into account quantum gravity  effects to all orders \cite{bm}. A systematic analysis of such states is almost completely lacking at present in string theory.

In the next sections of this paper, we provide a description in terms of open problems of various issues associated with infinities in the cosmological evolution of string-inspired  models.

\section{Tree-level string and M-theory effective cosmologies: The road ahead}
String effective actions have a scalar-tensor resemblance in zeroth-order $\alpha'$ expansion. Typical tree-level string actions involve, except for the dilaton and the graviton, a two-form potential in the NS-NS bosonic sector, a modulus field parametrizing the volume of the internal dimensions,  and a constant term related to the central charge deficit of string theory. There are a few known special exact solutions of such a theory describing  flat isotropic universes,
starting with the so-called linear dilaton background describing the first exact time-dependent solution of string theory \cite{lineardil1,lineardil2,lineardil3}.

However, in general, 
it is unknown whether these solutions are stable or not, or whether there exist singularities, or what happens to regimes where some of the fields involved blow up (cf.  \cite{gasp} and  therein). Long-term stability is an issue  of supreme  physical importance, and a lack of it implies that any model described by such an unstable exact solution with respect to perturbations is devoid of physical interest in the relevant regime. This is especially important for situations involving strong fields, asymptotic questions, and behaviours near singularities.

A well-studied example is the pre-big bang scenario solution  \cite{gv1}, and there are other important  exact solutions known, eg., the so-called dilaton-moduli-vacuum solution  \cite{cope},  belonging to the class of `rolling radii' solutions \cite{tsey}. There are also some very preliminary results on cyclic solutions \cite{bill,wands}, in addition to well-known solutions in the standard Brans-Dicke with $\Lambda$ (cf.  \cite{fujii}, chap. 4). There is also an instability result known in M-theory  \cite{bill2} (apart from the process of ekpyrosis, see below). The general area of string cosmology up to now is mainly concerned with  the study of effects that are based on these solutions.

However,  although there are by now a  number of exact homogeneous,
isotropic solutions in vacuum and with various matter fields in
scalar-tensor cosmologies and tree-level string models, questions of asymptotic stability such as the singularity problem are not yet
studied in such models with any serious degree of completeness.

Ideally one wishes to know the global stability behaviour of known exact string  cosmologies from every possible asymptotic point of view. Namely, a complete qualitative  classification of the various asymptotic  profiles including behaviour near singularities, the study of long-term asymptotics, and finally the question of genericity of the solutions, namely, their behaviour under suitable forms of perturbation.

We are particularly interested in classifying past singularities of string cosmologies
and their behaviour at early times, where all exact solutions are
dominated by a scalar field and so it is like having a vacuum.
It will also be useful to compare this with the analogous
situation in general relativity, perhaps using the conformal
equivalence between the string frame  dynamics and the Einstein frame
models.

We expect to have remarkable results here, both with respect
to  theorems giving sufficient and necessary conditions for singularity formation
in FRW or simple anisotropic  models, thus leading to a classification of the possible singularity forms (according to their Bel-Robinson energies etc), as well as with respect to the structure of infinity in these models, that is how the various fields decompose and balance near singularities, eventually to obtain detailed information as to  what their \emph{global} phase portraits look like (Poincar\'e method of central projection).

Armed with a classification of possible asymptotic forms, another important direction for future studies is  to include interacting fluids in the basic string actions (using both the viewpoint of the asymptotic splittings and that of the central projection
methods) in an effort to obtain valuable asymptotic information at early times, in
particular, in regimes where the scalar field that couples
nonminimally (as predicted by string theory) to the curvature dominates. This resembles the evolution in
vacuum, and it is particularly important that it is analyzed
thoroughly first. It will be very interesting to see how the various
known exact solutions fit into this scheme.

We further note that most scalar-tensor cosmologies utilize the standard coupling of the
scalar field $\phi$ to the curvature $R$ in the action, but in string theory
the scalar field is not universally coupled to all matter fields present in the
theory. However, the conformal transformation of these theories to
the Einstein frame usually introduces such a coupling, and the
matter lagrangian becomes a function of the scalar field $\phi$ as
well as the rest of the matter components, cf.  \cite{cot16}.

There are exact solutions corresponding to flat universes in this
case   \cite{cl-ba06b}, but due to non-geodesic motion
of test particles (due to possible violations of the equivalence
principle) one has to very carefully choose the couplings of $\phi$
to the matter, something which is a prerequisite in string theory. The interest here is in obtaining general results
about the early and late time attractors to these string cosmologies,
through the application of asymptotic methods, in an effort to
see whether or not, and in what sense, they tend to known limits.

One would also like to know the degree of genericity of the stable string cosmology solutions obtained from previous asymptotic analyses. An obvious first candidate to perturb is the pre-big bang solution  \cite{gasp}, or the cyclic universe  \cite{turok1}, or other singular solutions which are long suspected to be generic \cite{easther}.
The aim here is to use the degree of
genericity of the aforementioned exact solutions as a means to
decide about their physical significance. In general relativity, such
decisions have been made using experimental facts such as the
synthesis of light elements in the early universe and their
manifestations in the standard solar system tests of general
relativity  \cite{will}. In our string-theoretic framework, we propose the relative degree of genericity
of two solutions as a measure of their physical usefulness.

We expect the discussion in this section to naturally break in parts as follows:
\begin{itemize}
\item[A.] Classification of extreme states
\item[B.] Asymptotics of tree-level cosmologies
\item[C.] Asymptotics of universes in M-theory
\item[D.] Genericity of string cosmologies
\end{itemize}
We may further formulate specific problems related to the discussion so far.
\subsection{Asymptotics of specific tree-level string actions I}
The aim here is to describe the detailed asymptotic limits of universes obeying the basic gravi-dilaton effective string action, neglecting higher order and loop corrections. The plan to give a complete asymptotic analysis of the NS-NS models and first results about the asymptotic stability of the known solutions (dilaton-moduli-vacuum and axion solutions). This is a first necessary step in order to study the more advanced asymptotic analysis problems given below.

\subsection{ Asymptotics of specific tree-level string actions II}
The problem here is to study the asymptotic properties of universes with RR fields. This asymptotic analysis will tell us about the stability of the known solutions, whether they are unique. It is possible that there are solutions which exist only subdominantly asymptotically and these are expected to be identified by our analysis. This conjecture relates to a similar phenomenon previously observed in cosmologies with interacting fluids, namely, the so-called curvature exchange term entering subdominantly during the evolution to strong field states, cf. \cite{kittou1,kittou2}. This would constraint the types of singular asymptotic solutions.

\subsection{Asymptotics of specific tree-level string actions III}
The next step is to incorporate anisotropies into the asymptotic regimes and study simple anisotropic models with NS-NS and/or RR fields. There will be many decompositions of the vector field asymptotically and there may be solutions with a smaller number of arbitrary constants which still attract other families of solutions. We expect here the phenomenon of \emph{asymptotic cancellations }to occur, an effect first noticed in  \cite{kolionis2} (see also  \cite{kolionis3}). This will provide a detailed map of the solution space of string cosmologies.

\subsection{Asymptotics of M-theory cosmology}
It is very important that there are some exact cosmological  solutions in heterotic M-theory and  Horava-Witten cosmology (with nontrivial Ramond fields) found in the papers of Lucas-Ovrut-Waldram (cf.  \cite{low} and references therein), and so there are some basic regimes to perturb. An asymptotic analysis along the lines mentioned in this paper is expected to reveal for the first time the full significance of these solutions and especially the dimension of the asymptotic attractor they form.

\subsection{Asymptotics of M-theory cosmology II}
This is an extension of the previous problem in a different direction. The plan is to examine the stability of those cosmological solutions that are associated with a BPS state cf.  \cite{low1}. This requires the solutions to be inhomogeneous, and so may also have a significant role to play in the ambient cosmological  construction (cf. below).

\subsection{Genericity issues in string and M-theory cosmology}
For stable solutions found in the previous two problems, a genericity analysis to the full inhomogeneous perturbations may be applied. One expects in this way to discover what any generic solution sharing such characteristics will look like, at least for the analytic case. The genericity question in cosmology has a long history, starting with the studies of Khalatnikov-Lifshitz \cite{lk63,ll}, and continuing with Starobinski et al \cite{star1,star2,star3} for inflationary and other fluids. In higher-order gravity, a first analysis using Fuchsian series of arbitrary expansions for the metric was done in  \cite{bct} in the context of sudden singularity theory.
A complete analysis along these lines for a higher-order gravitational action in vacuum was performed in   \cite{trachilis1}.

\section{Universes with $\alpha'$ corrections}
Perturbing to the linear order the low-energy equations, string theory predicts the appearances of $\alpha'$ (higher-derivative) corrections modifying the effective action by adding  terms proportional to
\begin{equation}\label{alpha}
S_{\alpha'}\sim\alpha' \int du_g e^{-\phi}(R^2_{GB}-(\nabla\phi)^4),
\end{equation}
 where $R^2_{GB}=R^2-4Ric^2+Riem^2$ is the Gauss-Bonnet invariant\cite{art}. For an FRW metric, this has been recently  examined for asymptotic stability (without the dilaton term and coupling, but with curved universes allowed) as an $R+\alpha' R^2$ theory in the cases of radiation and vacuum in \cite{kolionis1,kolionis2}. The results in this context show that there are basically two attractors, the $t^{1/2} $ solution, and a Milne-type solution.

There are two `obvious' extensions of  these results to the following directions: First to include the dilaton term and couplings in the string effective action, and secondly, to do the asymptotic analysis  for any admissible value of the fluid parameter, $p=w\rho$. We note that already in vacuum, tracing the asymptotes was a highly nontrivial asymptotic problem, because one needed to choose the right variables for the vacuum to dominate asymptotically, cf. e.g.,  \cite{kolionis2}.

There are also a number of situations in string cosmology, for instance during the passage from pre- to post-big bang, that require, in addition to tree-level and $\alpha'$ terms, interaction matter terms\footnote{Such terms are also typically generated in the Einstein frame representation of a string effective theory that contains matter terms.}  \cite{gasp} as well as quantum loop correction terms \cite{ant1} in the string action. We can, however,  imitate some the effects of such terms as two or more interacting fluids present in the string action and  study the
asymptotics of solutions by adding  two interacting arbitrary fluids in the
theory (\ref{alpha}) that exchange energy at rates depending linearly on their
densities and expansion rate. Only partial results are known
\cite{chi02,ba-cl06b,cl-ba07,kittou1,kittou2} for the behaviour of particular exact
solutions in general relativity. Any results here will be the first
known ever to a similar situation in string cosmology. Useful tools for such qualitative and exact geometric asymptotics include  the method of asymptotic splittings and the dynamical system method of Poincar\'e compactification, suitable for these strong coupling corrections.

The various problems relevant to the marerial in this section  fall naturally in  the following three areas:
\begin{itemize}
\item[A.] Dilaton case asymptotics
\item[B.] General fluid
\item[C.] Interactions
\end{itemize}

\subsection{Asymptotics of gravi-dilaton flat FRW cosmologies}
The plan in this project is to extend relevant asymptotic analysis and stability results known separately in general relativity, higher-order gravity or pure scalar-tensor actions relating to  the flat and curved FRW universes, to the case of the full action (\ref{alpha}), that is including the dilaton term and its coupling to the higher-order curvature terms. This will provide especially important results for the stability of the vacuum in this effective string theory.
\subsection{Extensions to general tree-level cosmology for any fluid parameter}
For the basic string effective theory with quantum corrections, find all stable asymptotic solutions with a general fluid coupled to the dilaton. This involves several technical difficulties because of the nonlinearity of the equations, but we expect remarkable physical results to emerge from this project.
\subsection{Interacting fluids in flat string cosmology}
The aim of this project is to examine how all known single fluid results in flat string cosmologies generalize to the case of two interacting, more or less general fluids which exchange energy. This is very important because it will reveal which results remain in such a generalized situation, thus offering another clue to the physical viability of string cosmology. It will also compare with first results of interacting bulk fluids in brane theory as in \cite{ack} (see the work on mixtures in  \cite{ack}).
\subsection{Curved string cosmologies with interactions}
This deals with the important case of curved universes with interacting fluids in string cosmologies with higher order curvature corrections. We expect to see the effects of dominant as well as subdominant curvature asymptotically (cf.  \cite{kittou1} for the situation without a dilaton).
\subsection{Stability of 11-dimensional supergravity cosmology I}
The aim of this project is to see whether the properties of M-theory cosmologies studied in other projects above are stable with respect to perturbations formed when the M-theory action contains higher order corrections. We know that when the M-theory action is further deformed to include Lovelock and Weyl terms \cite{maeda1,maeda2}, there are several exact solutions not present in standard M-theory cosmologies. Are these solutions stable? In this problem one  expects to trace all possible asymptotic modes of the fields, and in this way to give the first reliable results about the possibility of a no-hair theorem for an inflationary stage in M-theory.
\subsection{Stability of 11-dimensional supergravity cosmology II}
Here the plan is to study the asymptotic stability of the deformed M-theory power-law cosmological solutions \cite{maeda1,maeda2}. This is especially important for the naturalness of inflation in M-theory cosmology.

\section{Braneworlds}
The possibility that our universe is described by a braneworld (brane in a large, higher dimensional bulk), motivated by the mass hierarchy within string theory, was instated in  \cite{aadd}.
There are at least two areas here where interesting research programs may be grounded. The first is the extension of the works \cite{kla1,kla2} (see next Section for a brief description of that work) from Minkowski (or dS or AdS) branes to general Robertson-Walker ones (for a background on the latter cf.  \cite{mann}, where embeddings, geodesics and fluctuations are worked out in detail). Since the equations of  an RW brane in a 5-dimensional bulk are  not envelopable by singular solutions, it is expected that they will in general contain regular asymptotic solutions. First promising results in this direction were presented in  \cite{kla3}, first for the scalar field case and then for more general fluids. In all cases, one needs to  perform detailed studies of asymptotic stability of the solutions using generalized asymptotic methods similar to those in \cite{kla3}. A stable, regular solution which would respect the energy conditions and localize gravity on the brane would be important and open the way to possible genericity questions, that is stability  under generic perturbations.

Secondly, it is very well-known that the existence of higher-order terms plays an important role in the process of ekpyrosis, in particular, in deciding about the asymptotic stability of the cyclic universe in string theory  \cite{gasp}. In the work  \cite{kolionis2} it was shown
that the existence of the Milne singularity and the attractor properties of our solutions bear a potential significance for the ekpyrotic scenario and its cyclic extension. In that regime, the passage through the singularity, `the linchpin of the cyclic picture', depends on the stability of a Milne-type state under various kinds of perturbations \cite{khouri1, khouri2,st1,st2}. In particular, during the brane collision, it is found that spacetime asymptotes to Milne and so it is expected that higher derivative corrections will be small during such a phase, cf. \cite{toley,erickson,leh}. The results of the  work  \cite{kolionis2} implies that such Milne states may indeed dynamically  emerge as stable asymptotes during the evolution, in any theory with higher-order corrections in vacuum or with a radiation content. What remains  is an interesting issue (that can be fully addressed with the asymptotic methods used in  \cite{kolionis2}) as follows: Find whether the `compactified Milne mod $\mathbb{Z}_2\times\mathbb{R}_3$ space'  monitoring the reversal phase in the ekpyrotic and cyclic scenarios, also emerges asymptotically as a stable attractor in the dynamics of higher-order gravity when the matter content is a fluid with a general equation of state.

The following research problems fall into categories as follows:
\begin{itemize}
\item[A.] Asymptotics of scalar field RW branes
\item[B.] Asymptotics perfect fluid RW branes
\item[C.] Genericity and stabiblity of ekpyrotic and cyclic scenarios
\end{itemize}

\subsection{RW branes and scalar fields}
The singularity structure and the corresponding asymptotic behavior of a 3D RW brane coupled to a scalar field in a five-dimensional bulk can be analyzed in full generality using the method of asymptotic splittings. One central issue is to examine the existence of regular solutions, and in accordance with the self-tuning proposal, to address the cosmological constant problem. Here the effects of curvature will play a role to see whether such solutions exist or whether the situation resembles that of a Minkowski brane which we referred to above.

\subsection{RW branes with perfect fluid}
This is a continuation of the previous problem to the case of the inclusion of a perfect fluid. This is a more general case which in the Einstein frame representation (meaning the brane action we considered previously), brane cosmology includes, for a special choice of the fluid parameter, the previous case of scalar field. New results in this direction are given in  \cite{kla4}. However, the issue of finding a regular bulk solution is further complicated by the existence of extra constraints, namely, how to localize gravity on the brane and at the same time meet the requirements of the null energy condition in the bulk. The content of the paper  \cite{kla4} shows that this is not possible for single fliuds, a result that points to our next problem.

\subsection{RW branes with interactions}
In this project we consider the standard gravi-dilaton string affective bulk action in the S-frame, in this case the vacuum is inequivalent to a $p=-\rho$ fluid and so we may have interactions. We may also take the Einstein frame representation of the theory and directly couple the scalar field with the fluid. This project will search the question as to whether regular asymptotics are the result of the interactions between the scalar field and the fluid. It would be interesting to test whether interacting fluids can meet all three conditions discussed in the previous problem.

\subsection{Genericity of the asymptotic solutions}
In this project the aim is to examine, through a generic perturbation analysis, whether the found solutions are stable when we consider general inhomogeneous perturbations. Through function counting, we may be able to conclude about the degree of generality of the found brane asymptotics. We note that the Landau-Lifshitz perturbation method discussed in previous sections can be applied to both regular as well as singular asymptotics.

\subsection{Ekpyrosis and the Milne state with asymptotically subdominant higher-order terms}
This project aims to examine the global asymptotic stability of the Milne state in string theories with higher-order corrections in an effort to decide as to whether or not this state is a viable representative of the passage through the singularity in models with an ekpyrotic or cyclic phase. In the prospective asymptotic analysis of this problem there will be hundreds of decompositions and dominant balances to consider one-by-one due to the combined effects of the general fluid and brane geometry. The plan is to start here covering first separately all those cases that have the relevant terms entering subdominantly. This will give a first indication of the possible stability. Subdominant evolution is more subtle than having all terms dominant, and it is known that it generally leads to surprising results.

\subsection{Ekpyrosis and the Milne state with asymptotically dominant higher-order terms}
This is a continuation of the previous problem concerning the asymptotic global stability of the Milne state in the Ekpyrotic and cyclic scenarios in M-theory. However, this time one is interested in the effects of the higher-order terms now entering dominantly in the evolution. This is the most nonlinear case. Upon completion, this project will bring new light not only to the process of ekpyrosis but one hopes to the whole of M-theory cosmology.

\section{Miscellanious results}
To appreciate the ambient construction and related issues developed in the following Sections, we outline here several distinct pieces of background, namely, various results from $AdS_5/CFT_4$ geometry, and more generally the ambient construction in conformal geometry, as well as various asymptotic limits of braneworlds. These results also have an independent  interest and importance in their own right.
\subsection{ $AdS_5/CFT_4$ geometry: The simplest ambient metric}
Witten in his work  \cite{w98} gives various proofs and implications of the following basic result: The $4-$dimensional Minkowski spacetime $\mathcal{M}_4$ is the boundary of $AdS_5$. A summary of the simplest properties associated with this construction  is given below in a sequence of steps.
\begin{itemize}
 \item The symmetry groups of bulk and brane agree on the boundary of bulk.
  \item Construct $(AdS_5,g_+)$ metric as the Poincar\'e (hyperbolic) metric on unit open ball $\mathcal{B}_5$ of $\mathbb{R}^5$ ($\sum_{i=0}^{4}y_i^2<1$ there), \beq g_+:=\frac{4|y|^2}{(1-|y|^2)^2},\quad |y|^2=\sum_{i=0}^{4}dy_i^2,\;y_0,y_1,\cdots,y_4\; \textrm{coordinates of}\;\mathbb{R}^5.\eeq
  \item $g_+$ does not extend everywhere on $\mathcal{B}_5\cup S^4$, because it is singular on the boundary $\partial\mathcal{B}_5=S^4$.
  \item Pick a function $\Omega=1-|y|^2>0$ on $\mathcal{B}_5$, and $\Omega=0$ on $S^4$.
  \item $g_+$ is conformal to a complete metric $\mathring{g}=\Omega^2g_+$ that extends smoothly on $\partial AdS_5=S^4$.
  \item $\mathring{g}|_{S^4}$ is a metric in $[g_4]$. The conformal infinity $\mathcal{I}_{AdS_5}=S^4$, that is its boundary.
  \item While $\mathcal{B}_5$ has a unique, well-defined metric, its boundary $\partial\mathcal{B}_5=S^4$ has only a conformal structure (both preserved under the actions of their  symmetry groups).
      \item Any function on $S^4$ extends uniquely to $AdS_5$ that has the given boundary values and satisfies the field equation.
      \item A conformal field theory on  $(S^4,[g_4])$ should be well-behaved.
      \item Maldacena conjecture: A string theory on $AdS_5\times S^5$ is equivalent to a certain SUSY Yang-Mills theory defined on $\mathcal{I}_{AdS_5}$.
      \item A black hole is then defined as a thermal state on the boundary, and the whole construction makes calculations easier because $S^4$ is conformally flat.
\end{itemize}
We conclude from this that one may proceed from the Poincar\'e metric $g_{AdS_5}=4(1-|y|^2)^{-2} g_E$,  to $\mathring{g}=\Omega^2 g_{AdS_5}$, and then restricting $\mathring{g}|_{S^4}$ to finally get a conformal structure on the boundary spacetime.

\subsection{The Fefferman-Graham fundamental theorem}
We now consider the \emph{inverse problem}: Starting from a conformal space-time manifold $(M,[g_4])$, is there a  metric on $V$ such that when we perform the construction we get the given conformal structure that we started with? This is the problem that occupied the fundamental work  \cite{fg}, concerned with the construction of conformal invariants. It was shown in   \cite{fg} that there exists a well-defined \emph{ambient metric} (this is the Fefferman-Graham metric) $g_+$ on $M\times\mathbb{R}$ (points $(x^\mu,y)$) with the following properties (cf.  \cite{fg}):
\begin{itemize}
  \item Locally around $M\times \{0\}$ in $M\times\mathbb{R}$, there is a smooth (non-unique) function $\Omega$ with $\Omega>0$ on $V$, $\Omega=0$ on $M$, and such that $\Omega^2g_+$ extends smoothly on $V$.
  \item $(\Omega^2g_{+})|_{TM}$ is non-degenerate on $M$ (that is its signature remains $(-++++)$ on $M$).
  \item $(\Omega^2g_{+})|_{TM}\in [g_4]$. ($M$ is the conformal infinity of $V$.)
  \item $g_+$ satisfies the Einstein equations with a cosmological constant $\Lambda$ to infinite order on $M$.
  \item $g_+$ is in normal form with respect to $g_4$:\[ g_+=y^{-2}(g_y+dy^2).\] Here, $g_y$ stands for a suitable formal power series with $g_0=g_4$. (We may also use $y$ as $\Omega$.)
  \item $g_+$ is unique: Given any two ambient metrics $g_+^1,g_+^2$ for  $(M,[g_4])$, their difference $g_+^1-g_+^2$ vanishes to infinite order everywhere along $M\times\{0\}$.
\end{itemize}
\subsection{Braneworld solutions, asymptotic limits}
As we discussed in previous sections of this paper, it is possible to have a complete profile of all asymptotic situations that emerge when we have a bulk 5-geometry $(V,g_5)$ containing an embedded 4-dimensional braneworld $(M,g_4)$ that is either a 4-dimensional  Minkowski, or de Sitter, or Anti-de Sitter spacetime, cf.  \cite{ack}. In general, all asymptotic solutions have a form dictated by the method of asymptotic splittings  \cite{skot}. This may be described in a series of steps:
\begin{itemize}
\item We have  bulk space $(V,g_5)$ (coordinates $A=(x^\mu,y)$) containing  embedded 4-dimensional braneworld $(M,g_4)$ (coordinates $x^\mu$, signature $(-+++)$)
filled with an analogue of perfect fluid $p(y)=\gamma\rho(y)$ and satisfying the 5-dimensional Einstein equations in the bulk, $G_{AB}=\kappa_5T_{AB}$.
\item We then assume the ansatz, $$ g_5=a^2(y)g_4+dy^2,\quad \textrm{for}  \; g_5\; \textrm{solutions on }\; V,$$
and look for solutions with $(M,g_4)$ being either a \textsc{Minkowski}, or \textsc{de Sitter}, or \textsc{Anti-de Sitter} spacetime.
\item In this case, the Einstein equations reduce to the generic form $\dot{x}=f(x)$, where $f$ is a smooth vector field and the solution vector is such that $x=(a,\dot{a},\rho)$
\item Solutions are then of the general form given by method of asymptotic splittings: $$a(y)=y^p\sum_{i=0}^{\infty}c_iy^{i/s},\quad y\rightarrow 0,\quad p\in\mathbb{Q},s\in\mathbb{N},c_i\in\mathbb{R},$$
and similarly for the density $\rho$.
  \item $g_5$ cannot be continued to arbitrary values in the $y$-dimension without some sort of matching; all flat brane solutions are singular at a finite, arbitrary $y-$distance from the position of the brane located at $y=0$. The generic curved problem is under investigation.
   \end{itemize}
From these results, it is not difficult to conclude that the following properties   apply in fact to a great variety of different models of braneworlds (cf. e.g.,  ~[\cite{rs1}]-[\cite{Forste2}] and  therein):
   \begin{itemize}
     \item The properties of the metric $g_4$ do not follow from those of the bulk metric $g_5$ but are dictated by field equations valid on the 4-`brane' itself.
  \item There is no conformal infinity for the 5-dimensional geometry (the brane is certainly  a kind of boundary to the bulk, but it can never be a \emph{conformal} boundary).
  \item No holographic interpretation is possible and there is no way to realize a boundary \textsc{CFT}.
\end{itemize}
In what follows, we present a novel approach in which all of the above difficulties are absent. For more details and developments, the reader is advised to look at ~[\cite{ac1,ac2}].

\section{Ambient cosmology}
We have discussed in previous Sections of this paper, various research problems which if studied will shed lights in tree-level string cosmology, in M-theory models of the universe, as well as in higher-order correction terms in the string action, and lastly in the extension to braneworlds. However, this is one more step to take and this is a further recent  extension of brane theory to a regime where all defaults mentioned above are absent.  This is a geometric construction we call \emph{ambient cosmology}.

In this Section, we present a brief summary of the main points of this construction to produce a situation where  generic spacetimes will end up having improved properties over those we may meet in the theory of hypersurfaces in general relativity (or in its higher-dimensional extensions as above). In the proposal below, a new bounding hypersurface, the conformal infinity  of a new cosmological metric in 5-dimensional `ambient' space will be the result.
One aspect of our results developed in  \cite{ac1,ac2} may be stated as follows:
\begin{theorem}
Let $(M,[g_4])$ be  a 4-dimensional spacetime with a conformal structure. Any  4-metric $g\in [g_4]$  has an ambient 5-metric $g_+$ on spacetime $V=M\times\mathbb{R}$ such that:
 \begin{itemize}
\item It satisfies the 5-dimensional Einstein equations with a fluid source on $V$
\item  $V$ has $M$ as its conformal infinity, $(\mathcal{I}_{g_{+}},\mathring{g}|_M)$
\item Any two conformally related 4-metrics on $M$, $g_1=\Omega^2g_2$,   have ambient metrics  differing  by $\mathring{g}_1|_M(0)-\mathring{g}_2|_M(0)=g_1$. Hence, $\mathcal{I}_{g_{+}}$ has a homothetic symmetry, $\mathring{g}|_M=cg_4$
 \end{itemize}
\end{theorem}

\subsection{The ambient cosmological metric}
Our construction is generally one belonging to conformal geometry (cf.  \cite{pen86}), and may be summarized as follows (cf.  \cite{ac1} for more discussion).
\begin{enumerate}
\item Take a 4-dimensional, non-degenerate `initial'  metric $g_{\textrm{\textsc{in}}}(x^\mu)$ on spacetime $M$. This step essentially involves the Penrose conformal method.
\item Conformally deform $g_{\textrm{\textsc{in}}}$ to a new metric $g_4=\Omega^2g_{\textrm{\textsc{in}}}$  by choosing a suitable conformal factor $\Omega$. This step connects the `bad' metric $g_{\textrm{\textsc{in}}}$ with the `nice', non-degenerate, and non-singular  metric $g_4(x^\mu)$.
\item Using the method of asymptotic splittings for the 5-dimensional Einstein equations with an arbitrary (with respect to the fluid parameter $\gamma$) fluid,  solve for the 5-dimensional metric $g_5=a^2(y)g_4+dy^2$ and the matter density  $\rho_5$.
\item Transform the solutions of step 3 to suitable factored forms of the general type, (divegent part) $\times$ (smooth part).
\item  Construct the `ambient' metric in normal form, $g_+$, for the 5-dimensional Einstein equations with a fluid. This is given by the following form,
\[g_+=w^{-n}\left(\sigma^2(w)g_4(x^\mu)+dw^2\right),\] $n\in\mathbb{Q}^+$, as $w\rightarrow 0$, with $\sigma(w)$ a smooth (infinitely differentiable) function such that $\sigma(0)$ is a nonzero constant.
\item $(M,[g_4])$ is the conformal infinity of $(V,g_+)$, that is   $\mathcal{I}=\partial V=M$.
\item The metric $g_+$ is conformally compact. This means that a suitable metric $\mathring{g}$ constructed from $g_+$ extends smoothly to $V$, and its restriction to $M$, $\mathring{g}|_M$, is non-degenerate (i.e., maintains the same signature also on $M$).
\item The conformal infinity $M$ of the ambient metric $g_+$ of any metric in the conformal class $[g_4]$ is controlled by the behaviour of a constant rescaling of the `nice' metric $g_4$.
\end{enumerate}
However, uniqueness of the ambient cosmological metric is not achieved like in the Fefferman-Graham construction  \cite{fg}. Instead,  we find [\cite{ac1}] an \emph{asymptotic condition} valid on the conformal infinity of the ambient space after taking suitable limits of the various possible geometric asymptotics of the problem. This method lies in the heart of the whole construction, and  is treated briefly in the next subsection.

\subsection{The asymptotic condition}
For any two conformally related 4-metrics $g_1,g_2$ in the conformal geometry of $M$, $g_1$ being the `good' (roughly meaning `regular') and $g_2$ the `bad' metric on the boundary, their ambient metrics $\mathring{g}_1|_M,\mathring{g}_2|_M$ differ by a homothetic transformation,
\beq\label{asymptotic}
\mathring{g}_2|_M(0)=c\,\mathring{g}_1|_M(0),\quad c:\;\textrm{const.}
\eeq

\textsc{Conclusion:} Starting from a conformal geometry on the spacetime $M$, the ambient cosmological metric returns a 4-geometry on  $M$ (its conformal infinity metric $\mathring{g}|_M$) that has a homothetic symmetry.

Therefore according to our proposal, as this is substantiated by the asymptotic condition, our 4-dimensional world is the conformal infinity of the ambient 5-space discussed above. What are the basic implications of this proposal? We may summarize some of them as follows.
\begin{enumerate}
\item\label{1} As a conformal manifold, $(M,[g_4])$ can have no singularities.
\item Cosmic censorship on $(M,[g_4])$ is equivalent to the validity of ambient 5-metric construction,  the asymptotic condition satisfied by  the ambient metric $\mathring{g}|_M$.
\item Global stability, asymptotic flatness.
\item Relation to $\mathbb{PN}$ twistor space.
\end{enumerate}
Below, we treat (\ref{1}) in some detail, and give a few  comments about  the rest towards the end of this paper.

\subsection{The Zeeman topology on the boundary}
Let us first state a celebrated result of C. Zeeman about the true topology of Minkowski space  \cite{z}.
\begin{theorem}
For Minkowski spacetime $M$, the group of homothetic symmetries (that is Lorentz transformations with dilatations) coincides with the group of all homeomorphisms of $M$ provided that its topology is not the usual Euclidean metric topology (that is $M$ is locally Euclidean) but a new one, called  the fine topology $\mathcal{Z}$.
\end{theorem}
The Zeeman topology has the following properties:
\begin{itemize}
  \item It is strictly finer than the Euclidean topology
  \item It possesses improved properties
  \item It extends to curved spacetimes
\end{itemize}
\textsc{Description:}
\begin{itemize}
\item For $x\in M$, an open ball in $\mathcal{Z}$ has the form $$B_{\mathcal{Z}}(x;r)= (B_{\mathcal{E}}(x;r)\setminus N(x))\cup {\{x\}},$$ where $B_{\mathcal{E}}(x;r)$ is the Euclidean-open ball, and $N(x)$ the null cone at $x$ (we remove  $N(x)$ and put back only the point $x$).
    \item Then $B_{\mathcal{Z}}(x;r)$ is $\mathcal{Z}$-open, but not $\mathcal{E}$-open.
    \item Hence, a set $A\subset M$ is $\mathcal{Z}$-open if $A\cap B$ as a subset of $B$ is $\mathcal{E}$-open, for every spacelike plane and timelike line $B$.
\end{itemize}
Zeeman also conjectured in  \cite{z} that an extension to the curved spacetimes of general relativity should be possible, that is for a general spacetime the homothetic group must be isomorphic to the homeomorphism group of the Zeeman topology, a conjecture  that  was shown to be correct by G\"{o}bel in  \cite{g}.

\subsection{Zeeman-G\"{o}bel theorem}
\begin{theorem}
For a general spacetime, the homothetic group is isomorphic to the group of homeomorphisms of the Zeeman topology.
\end{theorem}
We make the following comments.
\begin{itemize}
  \item Amongst all possible generalized topologies, the Zeeman topology is the unique one having this property, all others having homeomorphism groups isomorphic to the conformal group.
  \item For any spacetime $M$ in general relativity we have the freedom to choose either the standard Euclidean metric topology, giving $M$ the usual  manifold topology, or the Zeeman topology. It is of course the former that is used in all standard discussions of relativity.
  \item For our bounding spacetime $M$ - the conformal infinity of the ambient space $V$ - however, we do not have this freedom because we have shown the existence of a homothetic symmetry on $M$.
\end{itemize}

\subsection{Non-convergence of causal curves}
One notion that plays a key role in many theorems in global causal structure and the singularity thorems in general relativity is the convergence of a sequence of causal curves. Looking carefully at the proofs of various such results, we note the following (cf. ~[\cite{ac2}] for a more complete discussion of this).
\begin{itemize}
  \item For the convergence of a sequence of causal curves to a limit curve, one uses in an essential way the Euclidean balls with their Euclidean metric and their compactness in order to extract the necessary limits.
  \item Since the Zeeman topology is strictly finer than the Euclidean metric topology, such sequences will be Zeno sequences and their convergence in the Euclidean topology will not guarantee the existence of a limit curve in the Zeeman topology.

\end{itemize}
\subsection{Impossibility of singularities on $M$}
The non-convergence of sequences of causal curves has the following implication, cf. ~[\cite{ac2}].
\begin{itemize}
  \item In the proofs of the singularity theorems,  a contradiction appears when assuming the existence of a curve of length greater than some maximum starting from a spacelike Cauchy surface  $\Sigma$ (on which the mean curvature is negative) downwards to the past.
      \item One extracts a limit curve $\gamma$ (which locally maximizes the length between $\Sigma$ and an event $p$), and no curve can have length greater than that of $\gamma$.
      \item Here we cannot extract such a limit.
\end{itemize}
This result and also the more elaborate work  ~ [\cite{acp}] along these lines, opens the way for  the construction of complete spacetimes as the conformal infinities of physical theories in higher dimensional ambient space.

\subsection{Cosmic censorship}
We propose that the choice of metric in the conformal class $[g_4]$ ($g_4$ is the metric obtained after the conformal `cleaning' of the initial metric on $M$ (the latter is taken to satisfy the constraints of the Penrose conformal method),  must be made such that it does not spoil the non-degeneracy of the $\mathring{g}$ metric when restricted along the boundary $M$. The only way then left for which the five-dimensional ambient metric will lose its non-degeneracy on $M$ is when a timelike or null hypersurface forms somewhere in  $\mathring{g}|_M$, that is when there are naked points at infinity on the boundary spacetime. This would then make the ambient cosmological metric  $\mathring{g}|_M$ degenerate, contradicting the asymptotic condition. Therefore it seems that a choice must be made of those metrics $g_4$ in step 2 of the ambient procedure that respect cosmic censorship.

Conversely, the absence of naked singularities that follows from the validity of the asymptotic condition  on the ambient cosmological metric (in the sense of being valid on $\mathcal{I}_V$) has important implications, for it follows that a naked singularity may not be the end product of the process of Hawking evaporation of a black hole through thermal radiation. In this case,  future null infinity will generically meet the vertical line coming out of the spacelike singularity of the black hole due to the evaporation in the suitable Penrose diagrams, thus allowing material from inside the  spacelike singularity to be seen by an observer sitting at infinity. This is sometimes interpreted, as is well-known, as a possible violation of cosmic censorship at the quantum level. The deeper reason of why this works in the ambient framework we have developed is presently unknown.

\subsection{Global stability issues}
In this work, we have reviewed the idea (firstly advanced in  \cite{ac1}, [\cite{ac2}) that towards the Planck epoch classical spacetime becomes the conformal infinity of the ambient cosmological 5-metric, its conformal boundary. According to our proposal, spacetime achieves this by gradually (as we approach the Planck time) acquiring a conformal structure, this might correspond to some kind of conformal invariant - for instance the Weyl curvature - developed on $M$. In our construction, as we have already discussed, this is directly possible because of  the existence of one extra dimension,  the appearance of the ambient metric.

The key result here is the validity of the asymptotic condition (\ref{asymptotic}), instead of the usual uniqueness of the standard Fefferman-Graham ambient metric. This then is obviously related to some kind of stability of the original spacetime $M$ endowed with some metric away from the Planck time. For, any other metric $g'_4$  conformally related to  $g_4$ is a perturbation of  $g_4$  keeping the ambient construction unspoiled.

If, for instance, we are interested in the global stability of the Minkowski space, and consider some perturbation of it, then what we have shown here implies that we may replace it with any other, conformally related perturbation of it without disturbing the ambient construction. Therefore it seems that our construction is admitted by all conformally related perturbations of Minkowski space. This is something left to future work.

\subsection{Relation to twistors}
It is possible (although not clear at present) that our construction bears some relation to twistor space, in particular, the space $\mathbb{P}\mathbb{N}$ - the null projective twistor $5-$space.

There are also other known constructions in twistor theory, like the $\mathcal{H}$-space with cosmological constant, cf. \cite{le}, where a real $3-$manifold with a spacetime metric becomes the conformal infinity of another $4-$manifold that satisfies the self-dual Einstein equations with a cosmological constant $-1$.

In our case, we have a fluid in the 5-space, not a $\Lambda$, and the validity of the asymptotic condition is not so clear. Also the meaning of non-locality on our ambient space is not yet clear.

\section*{Acknowledgments}
One of us (S.C) is grateful to Professors R. Ruffini and R. Jantzen for making his presence in Rome during MG14 possible.

\end{document}